# Optical Coherence Tomography Angiography-OCTA dataset for the study of Diabetic Retinopathy

**Authors**

Pooja Bidwai [a], Shilpa Gite [a*], Biswajeet Pradhan [b*], Aditi Gupta [c], Kishore Pahuja [c]

**Affiliations**

[a] Symbiosis Centre for Applied Artificial Intelligence (SCAAI) Symbiosis Institute of Technology, Symbiosis International (Deemed University) (SIU), Lavale, Pune 412115 India

[b] Centre for Advanced Modelling and Geospatial Information Systems (CAMGIS), School of Civil and Environmental Engineering, Faculty of Engineering & IT, University of Technology Sydney, NSW 2007, Australia

[c] Natasha Eye Care Shiv Sai Lane Pimple Saudagar Pune Maharashtra 411027, India

**Abstract**

This study presents a dataset consisting of 268 retinal images from 179 individuals, including 133 left-eye and 135 right-eye images, collected from Natasha Eye Care and Research Institute in Pune, Maharashtra, India. The images were captured using a nonmydriatic Optical Coherence Tomography Angiography (OCTA) device, specifically the Optovue Avanti Edition machine as per the protocol mentioned in this paper. Two ophthalmologists then annotated the images. This dataset can be used by researchers and doctors to develop automated diagnostic tools for early detection of diabetic retinopathy (DR).

**Specifications Table**

| Topic | Eyecare |
|---|---|
| **Particular Topic** | Diabetic Retinopathy |
| **Format of data** | Excel file for grading and raw images |
| **Datatype** | JPEG photos with dimensions of 1596 x 990, 96 dpi, and OCTA (8x8 mm) |
| **Data Gathering** | OCTA scans are included in this dataset. Images of nonmydriatic OCTA were obtained with the Optovue Avanti Edition device. Three categories are used to categorize OCTA images: Moderate DR, Mild DR, and NO DR indications. |

| **Origin of Data** | Natasha Eye Care and Research Centre, Pune 411017, is where the dataset of 262 images was collected. |
|---|---|
| **Data availability** | Repository name: Optical Coherence Tomography Angiography Image dataset for detection of Diabetic Retinopathy. <br> DOI: https://doi.org/10.5281/zenodo.10400092 <br> URL : https://zenodo.org/record/10400092 <br> Under the terms of a data usage agreement, access to this dataset necessitates prior approval detailing the intended use of the data, along with the provision of the correct Digital Object Identifier (DOI) and the corresponding citation. |

## 1. VALUE OF THE DATA

The OCTA database is useful for the medical community as well as informatics researchers [1-3].

Here are the key points to note about the OCTA database in healthcare research:

1. For new researchers wanting to do early-stage DR research, the database is a great resource for education and research.

2. It's a foundation to develop and test automated diagnostic tools to detect early-stage DR so interventions can be done early and patient outcomes can be improved.

3. The database supports the training and testing of machine learning and deep learning models in retinal image analysis, facilitating the development of AI-enabled methods

4. By providing retina images annotated by experts, the database helps in formulating personalized treatment plans for DR patients considering their diagnosis [4,5].

## 2. BACKGROUND

Diabetic retinopathy (DR) is the leading cause of vision loss and blindness worldwide and in people with diabetes. It occurs when high blood sugar damages the blood vessels in the retina, causing swelling, leakage, and new abnormal blood vessel growth [6]. Early detection and treatment of DR can prevent severe vision loss. However, diagnosing DR, especially in its early stages, can be tricky because the retinal changes are subtle and varied [7]. That's why advanced imaging and artificial intelligence (AI) has been developed to improve diagnostic accuracy.

Optical Coherence Tomography Angiography (OCTA) is one such advanced imaging technology that has changed the way retinal diseases, including DR, are detected and monitored [8]. Unlike traditional imaging methods, OCTA provides detailed, high-resolution images of the retinal blood vessels without the need for dye injection. This non-invasive technique captures the microvascular structure of the retina, allowing clinicians to detect early signs of DR that may not be visible with other imaging modalities. OCTA, especially with devices like Optovue Avanti Edition, is becoming increasingly popular in clinics because it provides precise and detailed information about retinal health.

Despite the imaging technology advancements, there is still a need for automated diagnostic tools to help clinicians detect DR at its earliest stages. Machine learning and deep learning algorithms are showing great promise in this area, but they require large, well-annotated datasets [9-10]. These datasets provide the ground truth for training models to recognize the subtle signs of DR in retinal images. So, the creation of such datasets is the key to advancing AI in ophthalmology.

The dataset in this study consists of 268 retinal images from Natasha Eye Care and Research Institute in Pune, Maharashtra, India. It adds to the existing resources. The images are from both the left and right eyes of 179 individuals and annotated by experienced ophthalmologists. This dataset will help researchers and clinicians to develop automated tools for early DR detection and to improve patient outcomes through timely and accurate diagnosis.

## 3. DATA DESCRIPTION

Retinal OCTA images are useful for diagnosing many eye conditions. These images show anatomical structures of the eye like macula, optic disc, and blood vessels. Many retinal diseases can be diagnosed and treated by using the knowledge gained from these structures in retinal images [11-13].

DR is divided into two categories by Early Treatment Diabetic Retinopathy Study (ETDRS): Non-Proliferative Diabetic Retinopathy (NPDR) and Proliferative Diabetic Retinopathy (PDR). NPDR denotes the initial phase of the illness and is further classified into four categories: mild, moderate, severe, and extremely severe. PDR, on the other hand, is a more sophisticated version that falls into two stages: early and advanced. Based on the pathological disorders linked to diabetic retinopathy, medical professionals have divided the dataset into three categories. As a result, the dataset is categorized into three illness types: Moderate NPDR, Mild (or early) NPDR, and No DR symptoms. A file named "Annotations of the classifications.xlsx" contains the associated labels and classification status for 268 OCTA photos of 179 individuals, some of whom have diabetes and some of whom do not. This new dataset.

**Table 1.** The XLSX file's label representation for DR.

| Person | Image number | Image format | Classification Status |
|---|---|---|---|
| 1 | L1.1 | JPEG | NODR |
| 2 | L2.1 | JPEG | NODR |
| 3 | R3.1 | JPEG | NODR |
| 4 | L4.1 | JPEG | MILD_DR |
|   | R4.3 | JPEG | MILD_DR |
| 5 | L5.1 | JPEG | NODR |
| 6 | R6.1 | JPEG | MODERATE DR |
| 7 | L7.1 | JPEG | NODR |
| 8 | L8.1 | JPEG | NODR |
| 9 | R9.1 | JPEG | NODR |
| 10 | L10.1 | JPEG | NODR |
| 11 | L11.1 | JPEG | MODERATE DR |
|   | R11.3 | JPEG | MODERATE DR |
| 12 | L12.1 | JPEG | NODR |
|   | R12.3 | JPEG | NODR |

**Table 2.** Categorization of Database Images

| Classification | Number of images |
|---|---|
| NO DR | 226 |
| MILD DR | 31 |
| MODERATE DR | 10 |

Table 1 delineates the structured data contained within the XLSX document, providing the following definitions for its columns:

A. The attribute labelled "Person" refers to the identification number assigned to each patient within the dataset.

B. The "Image number" attribute denotes the total count of images associated with each patient, where "L" signifies an image of the left eye, and "R" represents an image of the right eye.

C. "Image Format" specifies the digital format of the images, which is .jpg.

D. "Classification Status" identifies the diagnostic categorization of DR for each case.

Furthermore, Table 2 elucidates the classification framework utilized within the database, highlighting the distribution of OCTA images across various categories.

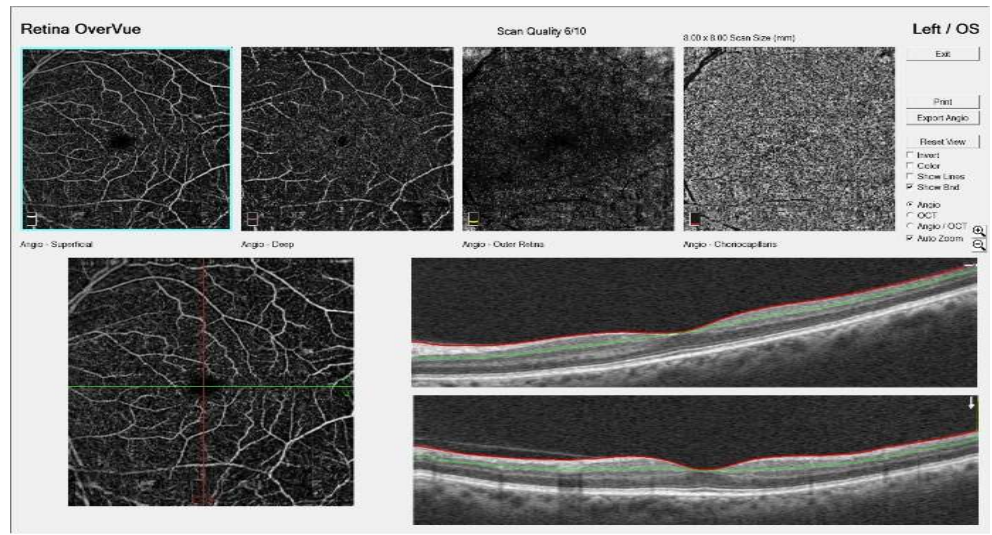

**Fig. 1a.** OCTA scan with no evidence of DR.

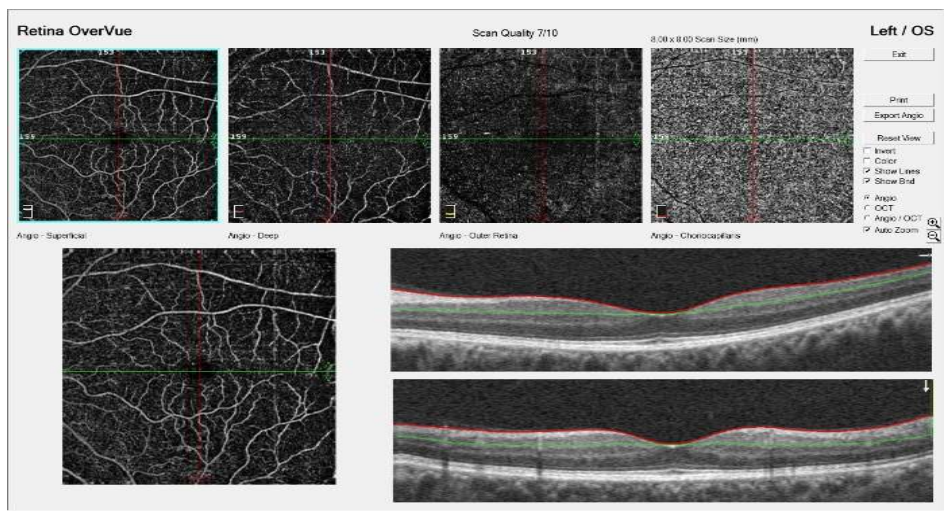

**Fig. 1b.** OCTA scan with Mild DR.

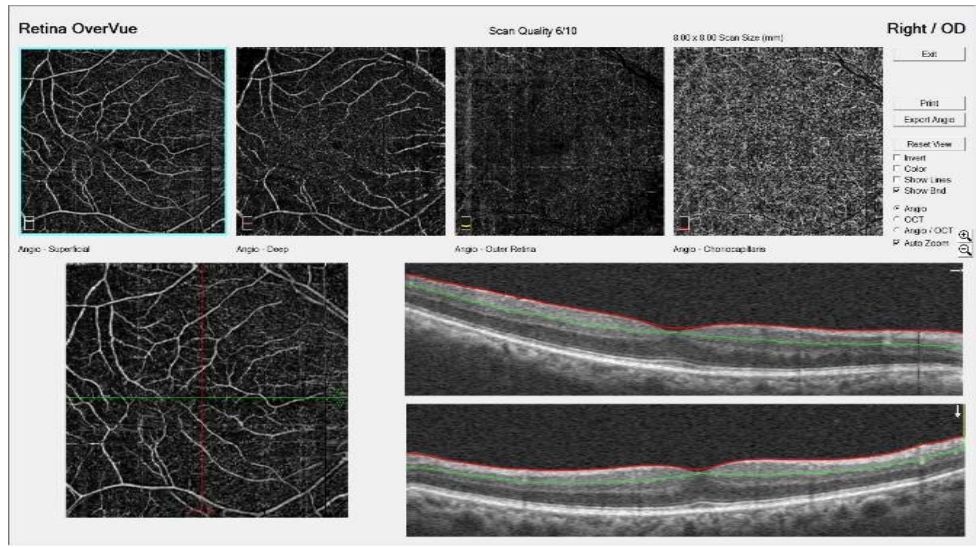

**Fig. 1c.** OCTA scan with Moderate DR.

Fig. 1a represents an OCTA image exhibiting no indications of diabetic retinopathy (DR). There are no detected microaneurysms or other related lesions in this instance.

Figure 1b depicts an OCTA image characterized by mild diabetic retinopathy. In this scenario, a minimal number of microaneurysms are observable.

Figure 1c showcases OCTA imaging of the individual, illustrating moderate diabetic retinopathy. This condition is marked by the presence of microaneurysms as well as retinal haemorrhages or exudates.

## 5. EXPERIMENTAL DESIGN, MATERIALS AND METHODS

### 5.1 Data Collection

In this research, a collection of 268 optical coherence tomography angiography (OCTA) images was acquired from individuals aged 30 years and older using a defined methodology. The cohort consisted of cataract patients within this age range, among whom a subset was diagnosed with diabetes while others were not. The study also included patients who had been living with diabetes for an extended period. On the 15th day post cataract operation, during the secondary follow-up visit, these participants' OCTA scans were taken at the Natasha Eye Care and Research Institute in Pune. The protocol is as follows:

- Participants were interviewed to collect demographic and medical history, and this was recorded in a patient file. This file had a checklist to capture information such as age, diabetes history, and other relevant health conditions like hypertension, ischaemic heart disease (IHD), or any medications being taken, which could be oral medications, insulin, or a combination of both.

- In cases with advanced complexity, systemic factors were deemed significant, encompassing conditions such as nephropathy, neuropathy, retinopathy, cardiomyopathy, vasculopathy, gastroparesis, and dermatological complications. Fig.2 shows the clinical details taken from the patients

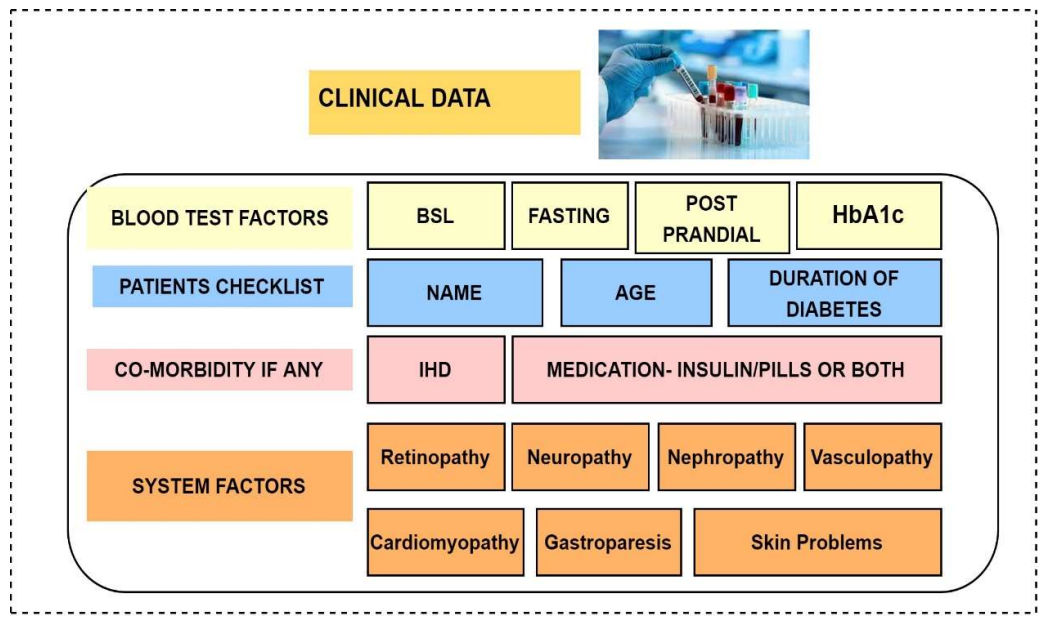

**Fig. 2** Patients Clinical Data

- OCTA images were taken with an 8x8 mm$^2$ wide scan centered on the macula by the optometrists in the hospital. These images covered the superficial capillary plexus (SCP), deep capillary plexus (DCP), and the full-thickness retinal slab as shown in Fig. 1. They were taken using the Optovue Avanti Edition OCTA Machine with the following specifications: Machine Model: RTVue XR 100 Avanti Edition, Field of View (FOV): 32o (H) x 22o (V), Detection System: Monochrome CCD Camera with WVGA 1/3o CCD Format, Near-Infrared (NIR) Illumination at 735 nm LED.
- Graded by a retina specialist, excluding images that were blurry, had artifacts, or had low image quality due to weak signal. Moreover, retinographies of patients who had undergone any other form of retinal treatment [14-15], Like PDR, post-laser therapy, etc, were systematically excluded from the study.

**5.2 Annotations**

In this study, a dataset consisting of 268 retinal images underwent a detailed annotation process by two experts, one ophthalmologist, and a retina surgeon. This process entailed the meticulous identification and mapping of retinal characteristics relevant to DR across each image. The annotations were executed with exceptional precision, enhancing the dataset's accuracy and dependability. This careful annotation by specialists bolsters the dataset's integrity, rendering it an exemplary basis for further investigations into the development of automated diagnostic tools for the early identification of DR.

**5.3 Exports**

The dataset, comprising 268 retinal photographs from 179 subjects, is made accessible for distribution to academics and medical professionals interested in advancing automatic detection systems. Each photograph, obtained through non-mydriatic Optical Coherence Tomography Angiography (OCTA) utilizing the Optovue Avanti Edition apparatus, includes precise labelling by experts highlighting the class of each image. This dataset is specifically intended to support the creation of sophisticated algorithms and machine learning frameworks, aiding the progress in enhancing early diagnosis and therapeutic outcomes for individuals with diabetic retinopathy.

## Limitations

- One of the biggest challenges is getting patients' consent and explaining the research. Furthermore, because an OCTA scan demands greater cooperation from the patient in order to get excellent clarity, it must be performed under the guidance of an expert. There are only 268 images in this study.

-Despite advancements in imaging technologies, there is still a significant need for more accurate and automated diagnostic tools to assist clinicians in detecting DR at its earliest stages [16]. Multimodal approaches, such as the fusion of Optical Coherence Tomography Angiography (OCTA) and Fundus images [17], have been developed to enhance the detection of DR. This multimodal image fusion technique has shown promise in detecting early signs of DR that may be missed by using either modality alone.

## Ethics Statement

The Symbiosis Institutional Ethics Committee (BHR) SIU granted authorization for the project under approval code SIU/IEC/583.

## Credit Author Statement

**Pooja Bidwai:** Conceptualization, Investigation, Data curation, Writing - original draft, Visualization; **Shilpa Gite:** Conceptualization, Investigation, Visualization, Review and Editing; **Aditi Gupta**: Investigation, Resources, Data curation, Review & Editing, Visualization; **Kishore Pahuja**: Conceptualization, Validation, Resources, Project administration; **Biswajeet Pradhan**: Validation, Review & Editing

## Data Availability

Repository name: Optical Coherence Tomography Angiography Image dataset for detection of Diabetic Retinopathy. URL: https://zenodo.org/record/10400092

## Declaration of Competing Interests

The authors declare that they have no known competing financial interests or personal relationships that could have appeared to influence the work reported in this paper.